\documentclass[english,aps,reprint]{revtex4}
\usepackage[T1]{fontenc}
\usepackage[latin9]{inputenc}
\usepackage{babel}
\usepackage{textcomp}
\usepackage{amstext}
\usepackage{graphicx}
\usepackage{subscript}
\usepackage[unicode=true]
 {hyperref}

\makeatletter
\@ifundefined{textcolor}{}
{%
 \definecolor{BLACK}{gray}{0}
 \definecolor{WHITE}{gray}{1}
 \definecolor{RED}{rgb}{1,0,0}
 \definecolor{GREEN}{rgb}{0,1,0}
 \definecolor{BLUE}{rgb}{0,0,1}
 \definecolor{CYAN}{cmyk}{1,0,0,0}
 \definecolor{MAGENTA}{cmyk}{0,1,0,0}
 \definecolor{YELLOW}{cmyk}{0,0,1,0}
 }

\makeatother

\begin{document}

\title{Addressing and manipulation of individual hyperfine states in cold
trapped molecular ions and application to HD$^{+}$ frequency metrology}

\author{U. Bressel, A. Borodin, J. Shen, M. Hansen, I. Ernsting, S. Schiller}

\address{Institut für Experimentalphysik, Heinrich-Heine-Universität Düsseldorf,
Universitätsstr.~1, 40225 Düsseldorf, Germany}
\begin{abstract}
Advanced techniques for manipulation of internal states, standard
in atomic physics, are demonstrated for a charged molecular species
for the first time. We address individual hyperfine states of ro-vibrational
levels of a diatomic ion by optical excitation of individual hyperfine
transitions, and achieve controlled transfer of population into a
selected hyperfine state. We use molecular hydrogen ions (HD$^{+}$)
as a model system and employ a novel frequency-comb-based, continuous-wave
5 \textmu{}m laser spectrometer. The achieved spectral resolution
is the highest obtained so far in the optical domain on a molecular
ion species. As a consequence, we are also able to perform the most
precise test yet of the ab-initio theory of a molecule. 
\end{abstract}
\maketitle
\begin{figure}[b]
\centering{}\includegraphics[bb=160bp 20bp 742bp 540bp,clip,width=0.8\columnwidth]{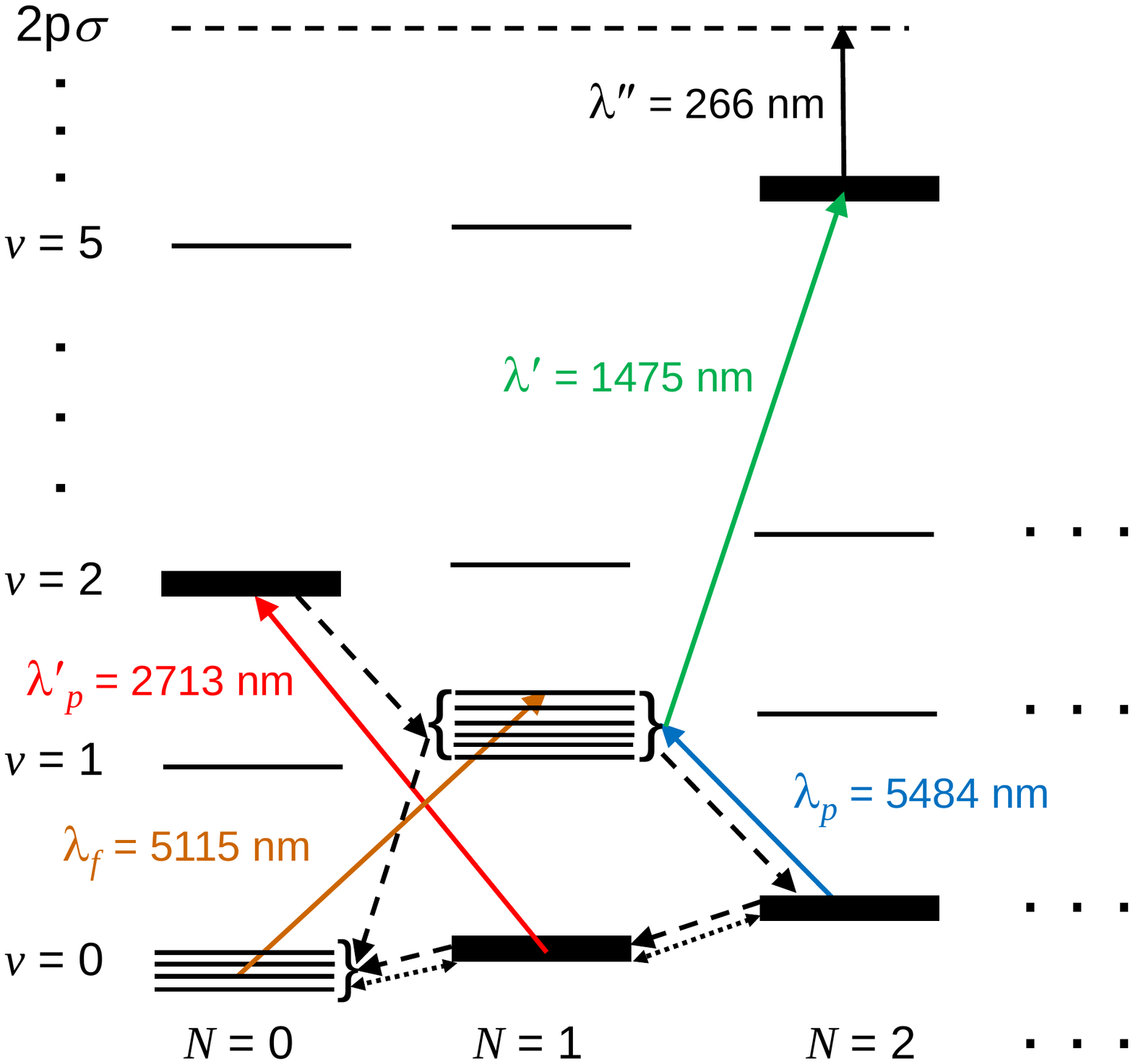}\vspace{-0.5cm}
\caption{\label{fig:Schematic-energy-level} Schematic energy level scheme
of HD$^{+}$ with transitions relevant to this work. Hyperfine structure
is shown schematically only for the $(v=0,\, N=0)$ and $(1,\,1)$
levels as lines, but is implied for all other levels as well (thick
bars). Full lines: laser-induced transitions, dashed lines: some relevant
spontaneous emission transitions; dotted lines: some black-body induced
transitions.  The spectrally narrow wave $\lambda_{f}$  selectively
excites molecules from a particular hyperfine state $(v=0,N=0,F,\, S,\, J)$
to a single hyperfine state $(1,\,1,\, F',\, S',\, J').$ Quantum
state preparation is performed by irradiating alternatingly the appropriately
tuned wave $\lambda_{f}\,$ and $\lambda_{p}\,$~, in conjunction
with spontaneous emission from the level $(1,\,1)$. Resonant laser
radiation at $\lambda'$  and nonresonant radiaton at $\lambda''$
 is used to detect that hyperfine-state-selective excitation to the
(1,1) level has occurred, by transferring the excited molecules to
the electronically excited molecular state 2p$\sigma$ from which
they dissociate. Initially, rotational cooling is performed by radiation
at $\lambda_{p}$ and $\lambda_{p}'$ . The level energy differences
are not to scale. The waves at $\lambda',\,\lambda'',\,\lambda_{p},\,\lambda_{p}'$
have large spectral linewidths and do not excite hyperfine state-selectively.}
\end{figure}

Cold trapped molecules \cite{Cold Molecules book,Cold Molecules book2}
currently represent an intense field of activity relying on sophisticated
methods of molecule production, translational and internal cooling,
spectroscopy and sensitive detection. Many applications, such as chemical
reaction studies \cite{Roth reactions,Drewsen group reactions}, tests
of molecular quantum theory \cite{Koelemeij 2007}, fundamental physics
\cite{Schiller and Korobov 2005,M=0000FCller 2004} and quantum computing
\cite{De Mille} would benefit strongly from the availability of advanced
manipulation techniques, already standard in atomic physics. These
are not straightforward for molecules, and for charged molecules have
not been demonstrated yet. Production methods for molecular ions (usually
by electron impact ionization) and, if heteronuclear, their interaction
with the black-body radiation of the surrounding vacuum chamber, usually
lead to significant population of a substantial number of internal
states. A first, important step in the manipulation of internal states
of molecular ions is population transfer between rotational states
(heteronuclear molecules usually being cold vibrationally, i.e. are
all in the $v=0$ ground vibrational state). It has been demonstrated
that a significant fraction (ca. 75\%) of an ensemble of diatomic
molecular ions can be pumped into the vibrational and rotational ground
level ($v=0,\, N=0$) \cite{Schneider 2009,Drewsen rotational cooling 2010},
see Fig. \ref{fig:Schematic-energy-level}. 

For a general diatomic molecule, however, this pumping is usually
not capable of preparing molecules in a single quantum state, because
spin interactions generate a hyperfine structure with several states
in each ro-vibrational level. For example, a diatomic molecule with
one unpaired electron ($s_e=1/2$), and nuclei with nuclear spins
$I_1=1/2,\,I_2=1$ (such as HD$^{+}$) has 4 hyperfine states in zero
magnetic field if the rotational angular momentum $N=0$, but 10 if
$N=1$, and 12 if $N\ge2$, see Fig. \ref{fig:Spin states of (0,0) and (1,1)}~a.
The ability to address selectively molecules in one particular hyperfine
state (or even in a single quantum state with a particular magnetic
quantum number $J_{z}$) and to transfer molecules from one hyperfine
state to another are clearly important tools of a molecular quantum
toolbox that can be part of a full quantum state preparation procedure. 

Complicating the addressing, the number of strong transitions between
two given ro-vibrational levels $(v,\, N),\,(v',\, N')$ is equal
to the larger of the two numbers of hyperfine states, i.e. potentially
high, and with only small differences in transition frequency. Fig.~\ref{fig:Spin states of (0,0) and (1,1)}~b
shows as an example the case of the fundamental vibrational transition
$(v=0,\, N=0)\rightarrow(v'=1,\, N'=1)$ in HD$^{+}$, where 10 strong
transitions occur over a range of about 60~MHz (\cite{Bakalov 2006,Bakalov 2010b}).
Addressing a single hyperfine state in a multi-spin molecule thus
requires a spectroscopy that can resolve individual {}``hyperfine''
lines in the spectrum. 

In this work, our approach is based on one-photon laser excitation
of the fundamental vibrational transition $(0,\,0)\rightarrow(1,\,1)$
at the wavelength $\lambda_{f}$, see Fig. \ref{fig:Schematic-energy-level}.
The relatively low transition frequency, $\lambda_f>2.5\,\mu$m for
diatomics, in combination with the low secular kinetic energy $k_{B}T_{sec}$
achievable by sympathetic cooling, yields a Doppler broadening $\Delta\nu_{D}$
of the transitions that is smaller than many line spacings. This provides
the desired quantum state selectivity for addressing some of the hyperfine
states, using strong transitions. For our test case HD$^{+}$, $\lambda_{f}=5.1\,\mu$m,
$T_{sec}\simeq10\,$mK, $\Delta\nu_{D}\simeq3\,$MHz. Additionally,
excitation of weak transitions (which violate the approximate selection
rules $\Delta F=0$, $\Delta S=0$, see Fig.~\ref{fig:Spin states of (0,0) and (1,1)}~b),
provides selectivity for all hyperfine states, since for these transitions
the frequency spacings are larger. Compared to the use of a pure rotational
excitation $(0,\,0)\rightarrow(0,\,1)$ or a microwave transition
within a ro-vibrational level, the use of a vibrational transition
has the advantage that the excitation may be followed by a much faster
spontaneous decay, either back into the ground ro-vibrational level
(rate approx. 6 s$^{-1}$, here) or into the (relatively long-lived)
rotational level $(v''=0,\, N''=2)$ (rate approx. 12 s$^{-1}$, here).
This allows reasonably rapid pumping of the molecule (possibly after
repeated absorption and spontaneous emission events) into another
long-lived state, a necessary condition for efficient quantum state
preparation, as shown below.

The experiment is performed on ensembles of HD$^{+}$ ions trapped
in a linear quadrupole radio-frequency trap (14.2~MHz), sympathetically
translationally cooled by co-trapped, laser-cooled Beryllium atomic
ions \cite{Blythe 2005} and rotationally cooled by lasers. Our laser
system consists of four subsystems: the $\lambda_{f}=5.1\,$\textmu{}m
laser spectrometer referenced to an atomic frequency standard \cite{Bressel}
(see Supplementary materials), a reliable, frequency-stabilized fiber-laser-based
313 nm laser for cooling of Beryllium ions \cite{Vasilyev 2011},
a rotational cooling laser system (a $\lambda_{p}=5.4\,$\textmu{}m
quantum cascade laser and, for part of the measurements, a $\lambda_{p}'=2.7\,$\textmu{}m
diode laser), and a pair of lasers $(\lambda,\,\lambda')$ for resonance-enhanced
multi-photon dissociation (REMPD), see Fig. \ref{fig:Schematic-energy-level}.
Rotational cooling \cite{Schneider 2009} is a crucial tool here,
as it significantly increases the fractional population of molecules
in the lower ro-vibrational level $(v=0,\, N=0)$, from ca. 10\% to
60 - 75\%. The possibility provided by our laser system to measure
the HD$^{+}$ fundamental vibrational transition frequency $f=c/\lambda_{f}$
also allows us to perform a precise comparison with the ab-initio
theory of the molecular hydrogen ion. 

Hyperfine-resolved ro-vibrational transitions are induced by the $\lambda_{f}=5.1\,$\textmu{}m
laser tuned precisely to individual transitions. In order to show
that this is actually achieved, we detect (without hyperfine state
selectivity) the population of the goal vibrational level ($v'=1,\, N'=1)$,
by 1+1' resonance-enhanced multi-photon dissociation (REMPD)\cite{Roth 2006},
see Fig. \ref{fig:Schematic-energy-level}. Together, this represents
a three-photon (1+1'+1'') REMPD process. The reduction of the HD$^{+}$
number by the REMPD process is our spectroscopic signal \cite{Roth 2006}. 

\begin{figure}
\begin{centering}
\includegraphics[width=1\columnwidth]{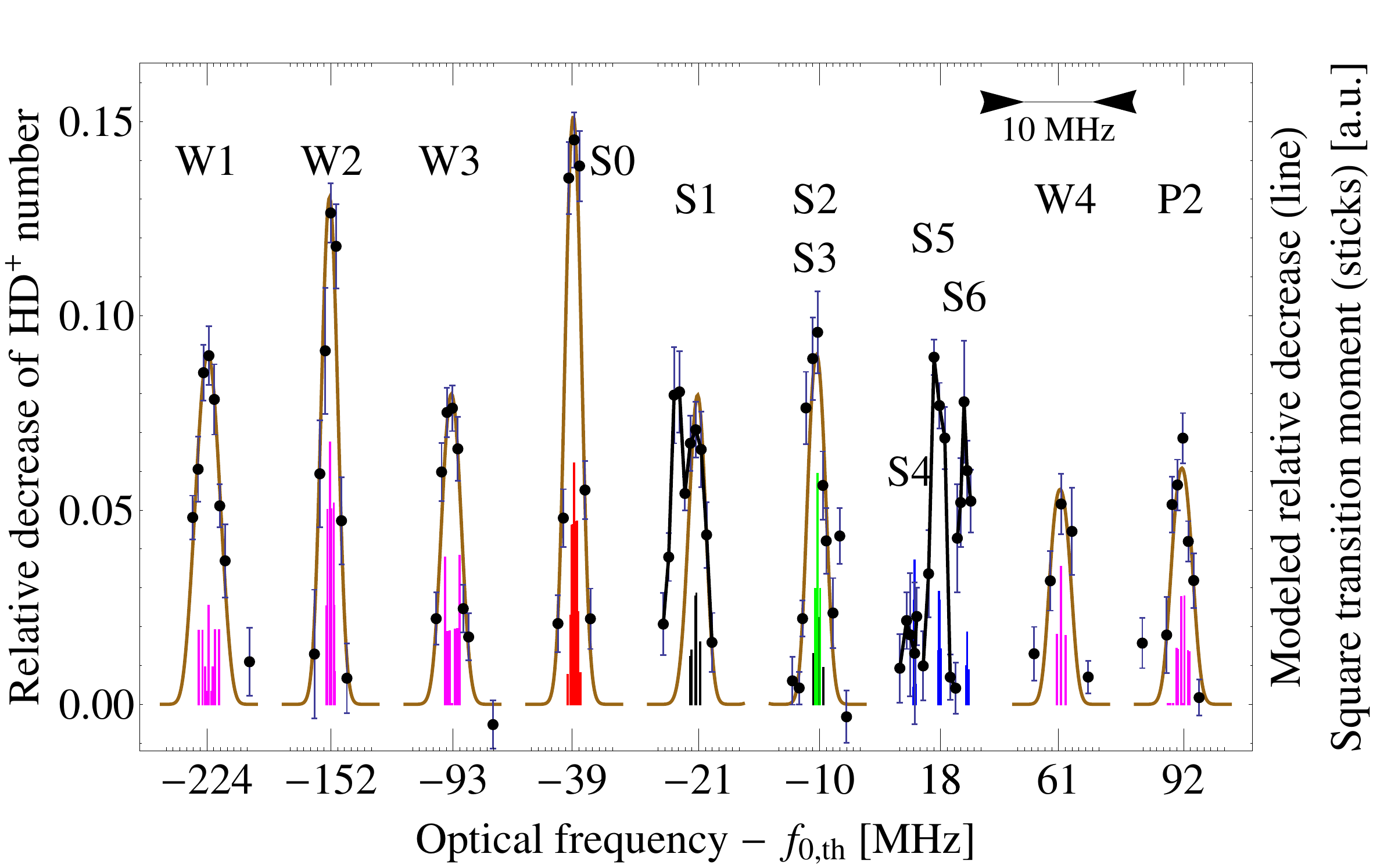}\vspace{-0.5cm}

\par\end{centering}

\centering{}\caption{\label{fig:Spectrum} Observed hyperfine spectrum of the \textbf{$(v=0,\, N=0)\rightarrow(1,\,1)$
}fundamental ro-vibrational transition in cold trapped HD$^{+}$ ions.
The effective intensity times irradiation duration product of the
5.1~\textmu{}m radiation varied from line to line, and was adapted
to avoid saturation. Brown lines are the result of fitting $f_{0,exp}$
and the individual line amplitudes for 9.5~mK temperature, and an
average magnetic field of 0.8~G. The sticks are for illustration
purpose and show the theoretical squared transition dipole moments
for the Zeeman components at 0.8 G, assuming exciting radiation polarized
at 45 degrees to the magnetic field. They are scaled by different
factors for presentation purpose. Color coding is as in Fig.~\ref{fig:Spin states of (0,0) and (1,1)}.
S4, S6 were taken at high intensity-irradiation time product, S5 at
a lower value. The side peak of S1 is probably due to an ion micromotion
sideband of S2/S3. The W4 line required hyperfine state optical pumping
for its detection (see Fig.~\ref{fig:Demonstration-of-spin state manipulation}). }
\end{figure}

Figure \ref{fig:Spectrum} shows the obtained hyperfine-state resolved
spectrum. All theoretically predicted and addressed hyperfine transitions
were observed; only the (nominally strong) transition S4 is barely
detected, for unknown reason. Transition W4, which originates from
a lower hyperfine state with only a small fractional population $(J=0$,
thus statistically containing only 5 - 6\% of all molecules), could
only be made clearly evident using a preceding hyperfine pumping step
(see below). Each of the 4 hyperfine states of the lower level was
selectively addressed, and 5 of the 12 upper level hyperfine states
were selectively populated. We also observed the line at $-10.2\,$MHz,
which contains two nearly coinciding transitions S2, S3, but originating
from different ground hyperfine states. The remaining strong transitions
(including the line marked P1 in Fig. \ref{fig:Spin states of (0,0) and (1,1)})
were also observed, but their small spacing prevents complete resolution,
and they are not reported in the Figure~\ref{fig:Spectrum}. 
\begin{figure}
\begin{centering}
\includegraphics[bb=7.5cm 1cm 18.300000000000001cm 500bp,clip,width=0.6\columnwidth]{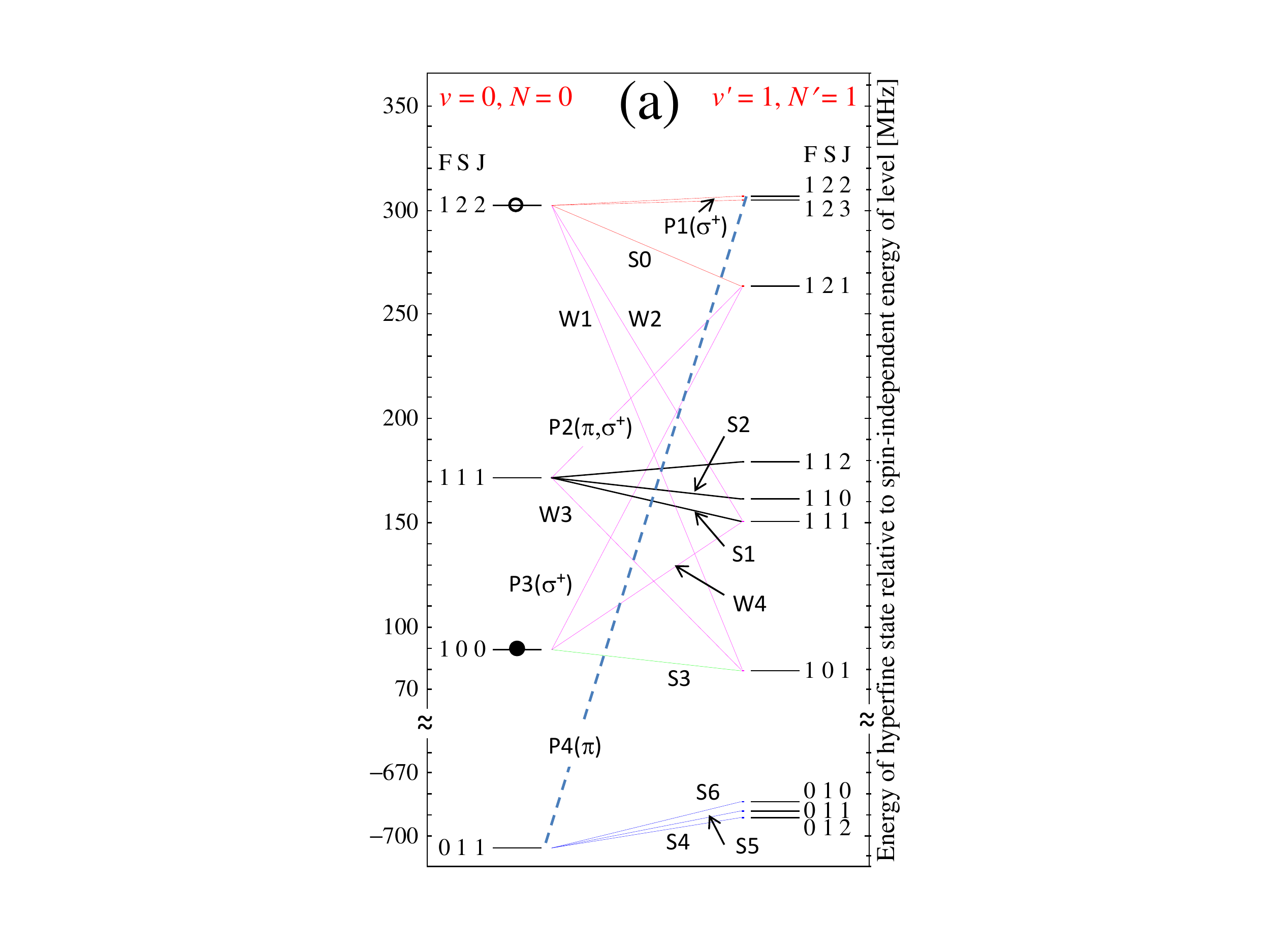}\includegraphics[bb=0bp 0bp 531bp 323bp,width=0.4\columnwidth]{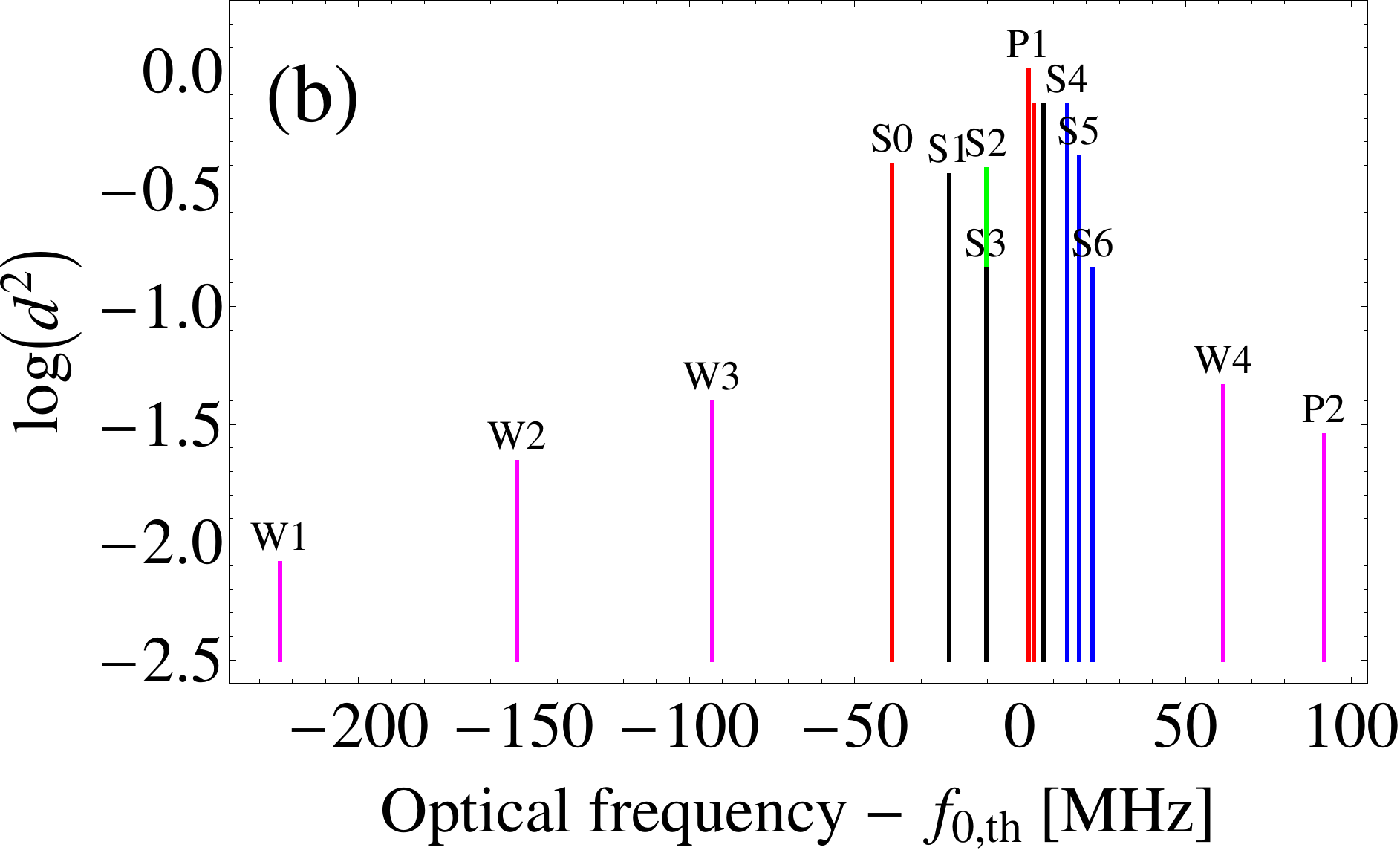}\vspace{-0.5cm}

\par\end{centering}

\centering{}\caption{\textbf{\label{fig:Spin states of (0,0) and (1,1)}}{\small a (left):
Energy diagram of the hyperfine states and main electric-dipole transitions
in zero magnetic field. b (right): Stick spectrum of the transitions
(in zero field; values of the squared transition moment $d^{2}$ are
normalized to the strongest transition). The states are labeled by
the quantum numbers $(F,\, S,\, J)$. Weak transitions are shown in
pink \cite{Bakalov 2010b}. Very weak transitions are not shown, except
for P4 (dashed). The {}``spin-independent'' transition frequency
$f_{0,th}$ is the value if nuclear and electron spin were zero. S0,
S1, S2, S3, S4, S5, S6, W1, W2, W3, W4, P2 are transitions studied
here ({}``W, S, P'' mean {}``weak'', {}``strong'', {}``pumping'',
respectively). All were observed except S4. W1, W3 are the transitions
used here to achieve population transfer from the hyperfine states
$(v=0,\, N=0,\, F=1,$ $\, S=2,\, J=2,\, J_{z})$ (empty circle) and
$(0,\,0,\,1,\,1,\,1,\, J_{z}')$ into the hyperfine state $(0,\,0,\,1,\,0,\,0,\, J_{z}'')$
(filled circle). P1, P2, P3, P4 are proposed optical pumping transitions
(with indicated polarizations) for preparation of the molecule in
the single quantum state $(0,\,0,\,1,\,2,\,2,\, J_{z}=+2)$ (one of
the Zeeman states in the open circled hyperfine state).}}
\end{figure}

\begin{figure}
\begin{centering}
\includegraphics[bb=0bp 0bp 480bp 295bp,clip,width=0.8\columnwidth]{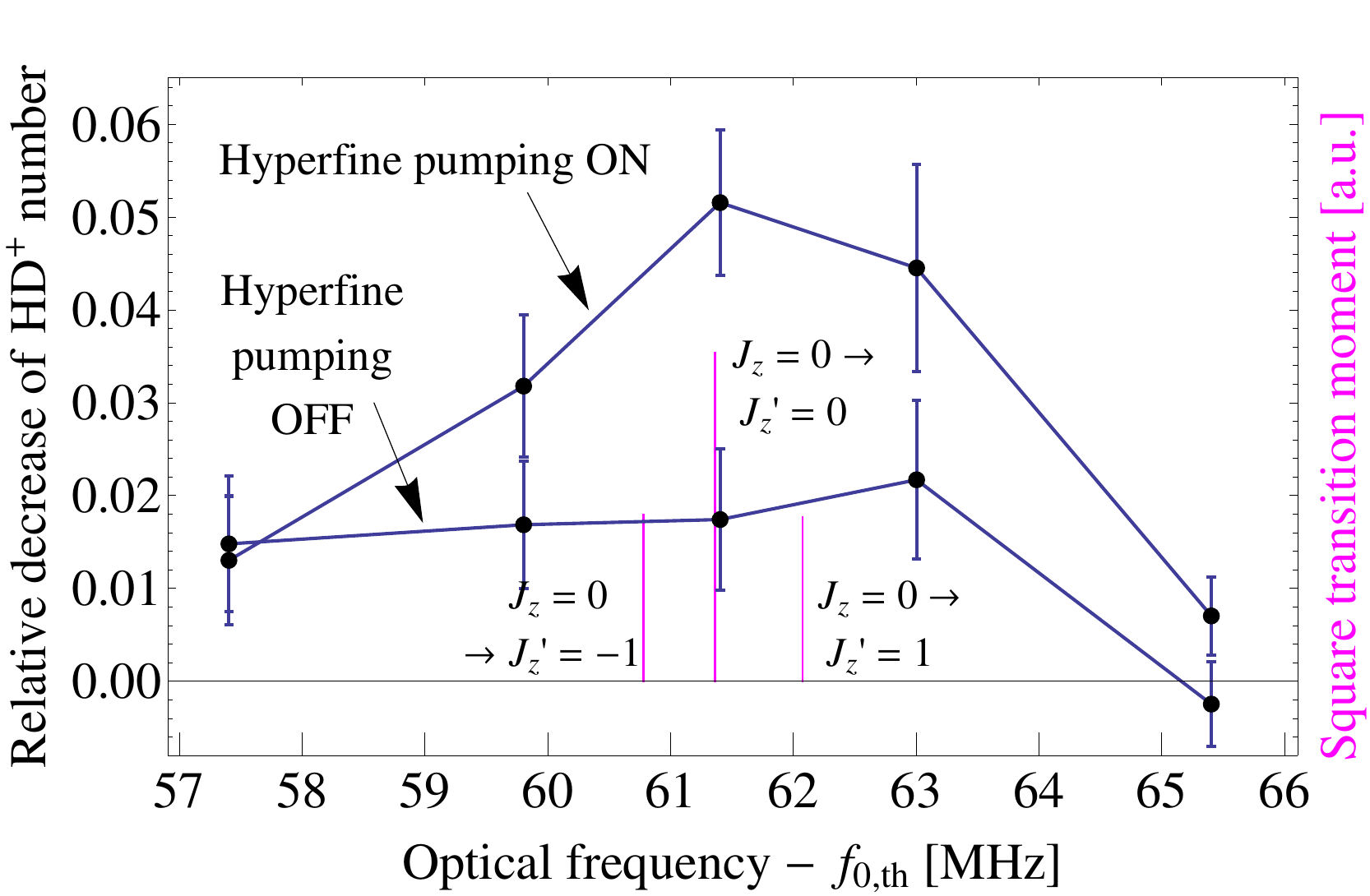}\vspace{-0.5cm}

\par\end{centering}

\centering{}\caption{\label{fig:Demonstration-of-spin state manipulation}Demonstration
of hyperfine state manipulation.  The transition W4 shown here is
observed only when hyperfine optical pumping is implemented. This
transition represents the excitation from a single quantum state,
$(0,\,0,\,1,\,0,\,0,\, J_{z}=0)$. Data shown was taken alternating
measurements preceded by hyperfine optical pumping (upper data points
joined by line) and not (lower data points). The intensity of the
5.1~\textmu{}m laser was set to its maximum both during hyperfine
pumping on the W1 and W3 transitions and subsequent detection of the
W4 transition. Irradiation time on the W4 transition was 3~s. Rotational
cooling by the 2.7~\textmu{}m and 5.5~\textmu{}m laser was used.
The zero level corresponds to the relative decrease measured when
the 5.1~\textmu{}m spectroscopy laser was blocked. The three sticks
show, for illustration purposes, the theoretical transition frequencies
and strengths in a 0.8~G magnetic field and radiation polarized at
45~degree to the magnetic field. The shift of the central component
is $-0.05$~MHz relative to the zero-field frequency.}
\end{figure}

We demonstrate hyperfine state manipulation by optical pumping of
individual hyperfine state populations into a goal state (filled circle
in Fig. \ref{fig:Spin states of (0,0) and (1,1)}). As a goal state
we choose $(v=0,\, N=0,\, F=1,\, S=0,\, J=0,\, J_{z}=0)$ which is
non-degenerate $(J=0)$ and thus a single quantum state. After rotational
cooling, we apply the following sequence twice: W1 line (3 s), rotational
re-pumping ($\lambda_{p}$ and $\lambda_{p}'$ simultaneously for
5 s), W3 line (3 s). A final 10 s of rotational re-pumping is performed
before the spectroscopic excitation. The W1 and W3 transitions excite
population from two initial hyperfine states $(0,\,0,\,1,\,2\,,2,\, J_{z})$,
$(0,\,0,\,1,\,1,\,1,\, J_{z}')$ (without $J_{z}$ selectivity) into
the same hyperfine state $(1,\,1,\,1,\,0,\,1,\, J_{z}'')$. This state
has dominant spontaneous decay to the goal state (green line S3 in
Fig.~\ref{fig:Spin states of (0,0) and (1,1)}). We find clear evidence
that this hyperfine state preparation is taking place by observing
the transition W4 starting from the goal state by REMPD, see Fig.~\ref{fig:Demonstration-of-spin state manipulation}.
This transition is not observable in our experiment without the preparation
procedure, since then the population in the lower hyperfine state
is too low. 

Our hyperfine-state resolved spectrum represents the highest-resolution
optical spectrum of any molecular ion so far \cite{Koelemeij 2007,Wing,Miller 1987}.
This enables an accurate comparison of experimental frequencies with
ab-initio theory. We obtain two hyperfine state separations in the
ground state, $\Delta f_{0,0,a}=(E(0,0,1,2,2)$$-E(0,0,1,1,1))/h$
and $\Delta f_{0,0,b}=(E(0,0,1,1,1)$$-E(0,0,1,0,0))/h$, from the
measured transition frequency combinations $f(S1)-f(W2)$, $f(W3)-f(W1)$,
$f(P2)-f(S0)$, and from $f(W4)-f(S1)$, $f(P3)-f(P2)$, respectively.
In addition, two excited state splittings, $\Delta f_{1,1,c}=(E(1,1,1,2,1)$$-E(1,1,1,1,1))/h$
and $\Delta f_{1,1,d}=(E(1,1,1,1,1)$$-E(1,1,1,0,1))/h$ are similarly
obtainable by suitable frequency combinations. A fit of these hyperfine
state separations to the data (fitting also the spin-independent frequency)
yields agreement with the ab-initio results $(\Delta f_{0,0,a},\,\Delta f_{0,0,b},\,\Delta f_{1,1,c},\,\Delta f_{1,1,d})_{th}$~$=$
$(130.60(1),$ $82.83(1),$ $113.33(1),$ $71.68(1))\,$MHz \cite{Bakalov 2006,Korobovprivcomm},
with deviations (\emph{exp.$\,-\,$theory}) of $(-0.22(0.13),$ $\,0.28(0.38),$
$\,-0.13(0.15),$ $0.27(0.19))\,$MHz. The two measured hyperfine
separations of the ground state also allow determining the two hyperfine
constants $E_{4}(0,0),\, E_{5}(0,0)\,$\cite{Bakalov 2006} which
fully describe the hyperfine structure of the ground state \cite{Bakalov 2010b}.
Our fit yields $(E_{4}(0,0),\, E_{5}(0,0))=$ $(906(17),\,$ $142.33(25))\,$MHz,
whereas the theory values are $(925.38(1),\,$ $142.29(1))\,$~MHz
\cite{Bakalov 2006,Korobovprivcomm}. 

Assuming instead that the hyperfine energies are given by the theoretical
values (this assumption being strenghtened by the agreement of hyperfine
theory and experiment for large-$v$ levels \cite{Bakalov 2006,Carrington}),
we can fit an overall frequency correction to the spectra W1, W2,
W3, S0, S1, S2+S3, W4, P2. We obtain the spin-independent frequency
$f_{0,exp}=$ 58 605 052.00 MHz, with combined statistical and systematic
error of 0.064 MHz (see Supplemental Materials). The theoretical value
is $f_{0,th}=$~58~605~052.139(11)(21)~MHz, where the first error
is due to the uncertainty of the fundamental constants and the second
is the theoretical error in the evaluation of the QED contributions
\cite{Korobov 2006,Korobov 2008,Korobovprivcomm}. The difference
between exprimental and theoretical resuls is $-2.0$ times the combined
theoretical plus experimental error. The relative experimental uncertainty
of $1.1\times10^{-9}$ represents the most accurate test of molecular
theory to date. In particular, our measurement is the first molecular
measurement sufficiently accurate to be explicitly sensitive to the
QED contributions of order $\alpha^{5}$ (relative to the nonrelativistic
contribution to the transition frequency), calculated as 0.109(21)
MHz for the transition studied here. 

Based on the technique demonstrated here, we can propose a realistic
optical pumping procedure for preparing most of the population in
a single quantum state $(v,\, N,\, F,\, S,\, J,\, J_{z})$, i.e. with
well-defined projection of the total angular momentum. Under typical
conditions, the relative statistical occupation of any individual
quantum state in $(v=0,\,N=0)$ is only $\simeq(1/12)\times(60\%-75\%)\simeq5\%-6\%$
under rotational cooling by a single laser $(\lambda_{p})$ or two
lasers $(\lambda_{p},\,\lambda_{p}')$. Exciting sequentially the
four transitions P4 $[(0,\,0,\,0,\,1,\,1)\rightarrow(1,\,1,\,1,\,2,\,2)]$,
P3, P2, P1$[(0,\,0,\,1,\,2,\,2)\rightarrow(1,\,1,\,1,\,2,\,3)]$ in
a weak magnetic field and with polarizations chosen as indicated in
Fig.~\ref{fig:Spin states of (0,0) and (1,1)}~a will cause transfer
of the population of all Zeeman quantum states of $(0,\,0)$ to the
single Zeeman quantum state ($0,\,0,\,1,\,2,\,2,\, J_{z}=+2)$, via
spontaneous emission processes from $(1,\,1,\,1,\,2,\, J'=\{1,\,2\},\, J_{z}')$,
which dominantly occur on strong transitions (red lines in Fig.~\ref{fig:Spin states of (0,0) and (1,1)}~a).
These excitations should be interleaved with rotational cooling (lasers
$\lambda_{p},\,\lambda_{p}'$), which also serves as repumper following
spontaneous decay into $(v''=0,\, N''=2)$. The optical pumping procedure
should take a few ten seconds and lead to 60\% - 70\% fractional population
in the goal state. 

In summary, we have shown that it is possible to address and prepare
individual hyperfine states in cold, trapped diatomic molecular ions
even in presence of a complex spin structure. A mid-infrared laser
spectrometer controlled by an atomic standard-referenced frequency
comb, and sufficiently low ion kinetic energies were two important
requirements. The observed, Doppler-limited, transition linewidths
(3~MHz) are the lowest obtained to date on a molecular ion species
in the optical domain (note that they scale as $(\hbox{\rm molecule\,\,mass})^{-1/2}$).
We also observed, for the first time to our knowledge, weakly allowed
hyperfine transitions using optical excitation. As one application,
we were able to directly determine the population fraction of molecules
in particular hyperfine states. The largest value we found was 19\%,
clearly indicating the effectiveness of our rotational cooling. We
also demonstrated excitation of a transition from a single quantum
state. Since our test molecule HD$^{+}$ is the simplest heteronuclear
molecule and is excited from the ro-vibrational ground state, this
study represents the first precision measurement of the most fundamental
electric-dipole allowed ro-vibrational transition of any molecule
\cite{Wing}. A comparison of theory with experiment showed that (i)
the hyperfine energies of small-$v,\, N$ ro-vibrational levels agree
within deviations of less than 0.3 MHz and (ii) the spin-independent
energy agrees within 2 times the error of $1.1\times10^{-9}$. 

\textit{ACKNOWLEDGEMENT}. We are grateful to B.~Roth, T.~Schneider,
A.~Yu.~Nevsky and S.~Vasilyev for help and support, and to V.~Korobov
for important discussions. Funding was provided by DFG project Schi
431/11-1.

\clearpage{}\pagebreak{}

\appendix

\section*{Supplemental Materials}

\subsection{General features of hyperfine structure}

We consider a diatomic molecule with total electron spin $s_e$, nuclei
with nuclear spins $I_{1},\: I_{2},\,$ in a rotational level $N$.
The spin degeneracy is lifted (only partially in zero magnetic field)
into a number of hyperfine states by the electron spin - nuclear spin
interaction ($\sim {\bf s\cdot I_i}$) and/or the electron spin -
rotation interaction ($\sim{\bf s\cdot N}$). In HD$^{+}$, the (approximate)
quantum numbers $(F,\, S,\, J)$ correspond to the couplings ${\bf F=s+I_p}$,
${\bf S=F+I_d}$, $\bf{J}={\bf S}+{\bf N}$. The electron spin - nuclear
spin interactions determine the main splittings in a given ro-vibrational
level, resulting in 4 singlets (if $N=0$) or in 4 multiplets (if
$N\ne0$). In the latter case, the electron-spin rotation interaction
determines the splitting strength within the multiplets \cite{Bakalov 2006}. 

The line splittings in the transition spectrum, Fig.~\ref{fig:Spin states of (0,0) and (1,1)}~b,
arise because in the lower and upper levels (i) the strengths of the
electron spin-nuclear spin interactions and of the electron spin -
rotation interaction differ, and (ii) the rotational angular momenta
$N$, $N'$ differ. The density of lines in the spectrum is essentially
independent of whether the transition is a fundamental vibrational
transition, an overtone transition ($\Delta v=v'-v>1$) (e.g. \cite{Roth 2006}),
or a pure rotational transition ($\Delta v=0$).

\subsection{5.1 \textmu{}m laser spectrometer }

The spectroscopy radiation was generated by frequency-mixing in a
nonlinear optical crystal two near-infrared lasers which are individually
referenced to a conventional frequency comb \cite{Bressel}. The generated
wavelength, 5.1 \textmu{}m, extends the coverge provided so far by
sources with absolute frequency measurement capability \cite{Firenze,Amy-Klein 2005}. 

We use two continuous-wave lasers, a Nd:YAG laser (1064~nm, ca. 6~W)
stabilized to a Doppler-free resonance in molecular iodine via its
second-harmonic wave at 532 nm, and a 1344 nm home-built external
cavity quantum-dot diode laser (QD-ECDL). The main part of the 1344
nm wave is resonantly enhanced in a ring resonator containing a periodically
poled MgO:LiNbO$_{3}$ crystal with poling period 25.4 \textmu{}m,
appropriate for difference frequency generation of 5.1 \textmu{}m
radiation. The 1064 nm wave is focused and overlapped with the 1344
nm wave in the crystal, but not resonated. Although the  crystal strongly
absorbs the generated 5.1 \textmu{}m radiation, an output power of
up to 0.1~mW is generated. Small fractions of the 1064 and 1344 nm
waves are sent via an unstabilized single-mode optical fiber to another
laboratory containing the femtosecond Ti:sapphire frequency comb,
where both frequencies are measured. The 1344 nm laser is actively
frequency-stabilized to a mode of the frequency comb. The 1064 nm
wave frequency is continuously measured by the comb. The comb itself
is stabilized to a hydrogen maser or ultra-stable quartz oscillator,
both referenced to GPS.

Tunable, frequency-stable 5.1 \textmu{}m radiation is produced by
amplitude modulating the diode laser output wave with an integrated-optic
modulator at a variable radio-frequency $\Omega$ before sending the
wave to the resonator. This generates two sidebands, and the resonator
length is stabilized to resonate one of the two, generating a single
frequency at 5.1 \textmu{}m. By changing the modulation frequency
$\Omega$, the 5.1 \textmu{}m radiation can be smoothly and precisely
frequency-tuned over 460~MHz. We determined the spectral width of
the 5.1 \textmu{}m radiation as $\Gamma_{f}\simeq0.68\,$MHz. The
frequency instability (Allan deviation) of this radiation is constant
at ca.~23 kHz for integration times $\tau$ up to 10 s, dropping
to 4 kHz at $\tau=800\,$s. Both linewidth and instability are significantly
lower than the Doppler width of the HD$^{+}$ hyperfine transitions
studied here. The absolute frequency uncertainty of the 5.1 \textmu{}m
radiation is less than 10 kHz when averaged over 1 h.

\subsection{Procedures and analysis}

\begin{figure}
\centering{}\includegraphics[width=1\columnwidth]{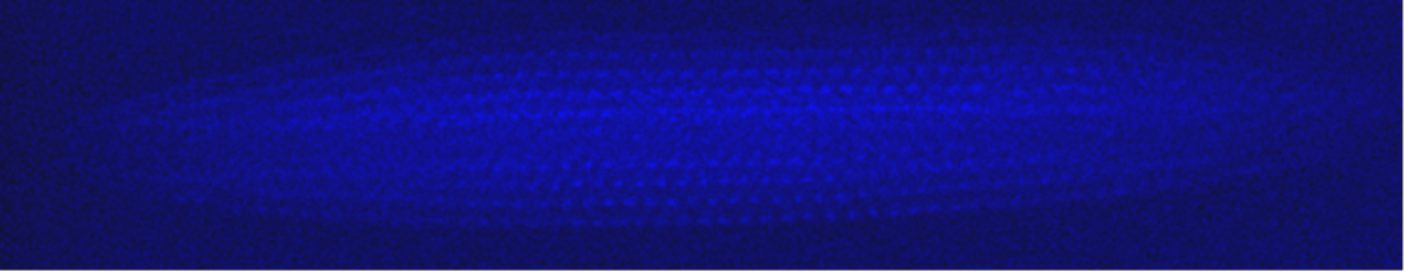}\caption{\label{fig:CCD-image-of}CCD image of a typical Beryllium ion Coulomb
cluster with embedded HD$^{+}$ ions (darker region along the long
axis), as used in the experiment.}
\end{figure}

Each data point in a spectrum measurement is acquired from one loading
of HD$^{+}$ ions into a laser-cooled Be$^{+}$ ion cluster. Fig.~\ref{fig:CCD-image-of}
shows an example.

From molecular dynamics (MD) simulation of ion cluster fluorescence
images \cite{Zhang 2007-1} we deduce an upper limit for the secular
temperature $T_{sec}\le15\,$mK. After loading, the molecules are
rotationally cooled by $\lambda_p\simeq5.5\,\mu$m radiation (and,
for some measurements, also $\lambda_{p}\text{\textasciiacute\,=2.7\,}\mu$m
radiation) for 30~s, transferring about 60\% (75\%) of the molecules
into the $(v=0,\,N=0)$ ro-vibrational ground level. This population
is spread over the four hyperfine states. Then, the rotational pumping
laser waves are blocked and the $5.1\,\mu$m spectroscopy radiation
is applied along the trap axis for 3 - 30 s, simultaneously with a
1475 nm diode laser (ca. 1~mW) driving the $(v'=1,\, N'=1)\rightarrow(v''=5,\, N''=2)$
overtone transition, and a 266 nm laser (ca. 25 mW) that dissociates
the molecules in the latter state. 

The number of trapped HD$^{+}$ ions is measured by excitation of
their secular motion before the rotational cooling and after the (partial)
photodissociation. Excitation of the secular motion heats up all ions
and therefore changes the Be$^{+}$ fluorescence rate, which is detected
by a photomultiplier. The relative decrease in the photomultiplier
signal is the signal of interest, the excited molecule fraction. As
the spectroscopy source linewidth is 0.7 MHz, we made frequency steps
of typically 0.8 MHz when scanning through the lines. One data point
could be recorded on average every 3 min. This limited the number
of data points per frequency value that could be taken. We performed
between 4 and 25 measurements per frequency point. Typically, one
hyperfine transition required one day of measurement. Background level
measurements were performed in the same way, with the 5.1 \textmu{}m
laser blocked. Alignment of the 5.1 \textmu{}m laser with the ion
cluster varied from day to day, which was compensated by adjusting
the power.

The typical standard deviation of the relative decay data taken at
a given optical frequency (not of the mean) is 0.02. This is due primarily
to the low number of HD$^{+}$ ions loaded into the ion cluster (a
few hundred) and the low relative population in a particular hyperfine
state (only a fraction $\simeq(0.6-0.75)$$\, p_{J}$, where $p_{J}=(2J+1)/12$
in a particular hyperfine state $J$, for $N=0$), which is further
reduced during the REMPD phase ($\lambda_{p},\,\lambda_{p}\text{\textasciiacute}$
lasers off) by the competing black-body-radiation driven excitation
$(v=0,\, N=0)\rightarrow(0,\,1)$. The overall small numbers lead
to significant statistical fluctuations in the number of ions actually
prepared into a particular initial hyperfine state after optical pumping,
and in the fraction excited and dissociated. As a consistency check,
we found that the maximum observed dissociated fractions (at high
intensity and on resonance) did not exceed $p_{J}$. For example,
for line S0, we observed $(19\text{.4\ensuremath{\pm}}1.4)$\% ion
number reduction, for lines S2+S3, $(16\text{.4\ensuremath{\pm}}2.8)$\%.
Note that the values shown in Fig.~\ref{fig:Spectrum}, are lower
because there, lower intensities were used in order to avoid saturation
broadening.

The smallest observed full-width-half-maximum linewidths are $\simeq$~3.0
MHz. Assigning this to be due to Doppler broadening only, the value
yields an upper limit of 15 mK for the secular temperature, consistent
with the MD simulations. 

The observable spectrum may be modeled, including the effects of finite
secular temperature, finite laser linewidth, finite population in
the lower level, the effect of unresolved Zeeman splittings by the
non-zero magnetic field in the ions' region \cite{Bakalov 2010b},
and black-body-radiation induced excitation to (0, 1). In order to
simplify the analysis, in the experiment we chose laser power and
irradiation times such that significant saturation and concomitant
broadening of the signals was avoided, i.e. the linear absorption
regime was maintained. We furthermore take into account: (i) from
independent measurements we have some knowledge about the magnetic
field: an upper limit of ca. 1 G \cite{Shen}; (ii) the laser linewidth
does not contribute strongly to the total linewidth, and we may therefore
use an effective Doppler temperature for modeling the lineshape; (iii)
the overlap between the focused spectroscopy beam (ca. 0.5 mm waist)
and the ion cloud was not constant over the time span covering all
hyperfine line measurements; (iv) the effect of black-body-induced
excitation $(0,\,0)\rightarrow(0,\,1)$ represents a reduction in
maximum observable signal. Thus, we fit an effective product of intensity
and irradiation time to each transition spectrum in order to reproduce
the signal levels. We used a simplified model of the magnetic field
inhomogeneity in the ions' volume. We find a good fit for an average
magnetic field of 0.8 G and $(9.5\pm1)$ mK temperature. The statistical
error for the spinless frequency $f_{0,th}$ is obtained from a Monte
Carlo simulation as 60 kHz.

\subsection{Systematic errors}

Due to the slow data rate of this experiment, measurements of systematics
were not possible. However, the well-developed theory of the HD$^{+}$
molecule allows relying on theoretical results to estimate upper limits
for various systematic errors. Those due to light shift, black-body
shift \cite{Karr 2006}, electric quadrupole shift \cite{Quadrupole moment},
and Stark shift \cite{Moss 2002,Baklov 2011}, are theoretically estimated
to be less than 10~kHz. The only potential significant effect is
the Zeeman effect. The sticks in Fig. \ref{fig:Spectrum},~\ref{fig:Demonstration-of-spin state manipulation}
show the splittings expected for the typical magnetic field value
in our trap \cite{Bakalov 2010b}. These unresolved splittings could
lead to a shift of the centers of the lines but we find that the weighted
mean frequency of the magnetic components of any line shifts weakly
with magnetic field (e.g.~less than ca. 10~kHz for line S0 at 1~G).
We modeled the influence of the imprecisely known magnetic field on
the fitted spin-independent frequency and estimate an error of 13~kHz
from this effect. In order to obtain the spin-independent frequency
from our measurements, we use the theoretical prediction of the hyperfine
energies. The influence of their theoretical errors onto the fitted
spin-independent frequency is conservatively assumed to be 10~kHz.
The error of the optical frequency measurements is less than 10~kHz.
Similar considerations are applied to the fit of the hyperfine splittings.

\subsection{Theoretical transition frequencies}

The theoretical frequency of any particular hyperfine transition arises
from three contributions. The main contribution is non-relativistic,
58 604 301.269 MHz, determined by the solution of the Schrödinger
equation \cite{Korobov 2006} and whose error (11~kHz) comes dominantly
from the uncertainties of the electron-proton and proton-deuteron
mass ratios $m_e/m_p$, and $m_p/m_d$ \cite{CODATA2010}. A second
contribution, 750.870~MHz, is from relativistic and QED effects \cite{Korobov 2008},
with estimated theoretical uncertainty of 21~kHz \cite{Korobovprivcomm}.
The sum of these two contributions is the spin-independent frequency
$f_{0,th}$. The third is the hyperfine contribution \cite{Bakalov 2006},
e.g. -38.685~MHz for transition S0. Improved values for the hyperfine
constants and Bethe logarithm communicated recently by V. Korobov
have been used here. The influence of these improved values as compared
to the published ones on the results presented here is small in comparison
with the experimental uncertainties. The theoretical uncertainty of
the hyperfine contributions to the transition frequencies is less
than 10~kHz \cite{Korobovprivcomm}.

\subsection{Hyperfine state preparation}

The re-pumping suggested acts predominantly on the strong, non-spin-state
changing transitions. P2 is a weak transition, P3, P4 are very weak
transitions. The transition dipole moments of the latter are so small
\cite{Bakalov 2010b} that a quantum cascade laser (typical output
power level 10 mW) could be used for obtaining usefully large rates.
Such a laser could be frequency-stabilized to a spectrometer of the
type developed here, using the beat note with it as an error signal.

An additional rotational cooling laser that depletes the $(v=0,\, L=3)$
level by exciting from it to the $(1,\,2)$ or $(2,\,2)$ level could
enable an even higher population fraction in the goal hyperfine state. 

In a similar way, any other Zeeman states of the ground vibrational
level could in principle be populated to a high fraction. 

It is recognized that variations of the above scheme can also be used
to achieve similar population fractions in individual quantum states
of the $(v=0,\,N=1)$ level.

\end{document}